\documentclass[aps,prd,twocolumn,preprintnumbers,superscriptaddress,dblfloatfix,nofootinbib]{revtex4-1}
\usepackage[utf8]{inputenc}
\usepackage{lmodern}
\usepackage{amsmath,amssymb,mhchem}
\usepackage{graphicx}
\usepackage{url}
\usepackage{color}
\usepackage{enumitem}
\usepackage{subfigure}
\usepackage[dvipsnames]{xcolor}
\usepackage[colorlinks=true,breaklinks=true]{hyperref}
\hypersetup{allcolors=[rgb]{0.0 0.0 0.6},linkcolor=[rgb]{0.75 0.05 0.05}}
\usepackage{bm}
\usepackage{epsfig}
\usepackage{amsmath}
\usepackage{amssymb}
\usepackage{slashed}
\usepackage{color}
\usepackage{accents}
\usepackage{url}
\usepackage[dvipsnames]{xcolor}

\newcommand{\exclude}[1]{}

\begin{document}

\preprint{IPMU21-0080}
\preprint{RIKEN-iTHEMS-Report-21} 

\title{Impacts of Jets and Winds From Primordial Black Holes}

\author{Volodymyr Takhistov}\email{volodymyr.takhistov@ipmu.jp}
\affiliation{Kavli Institute for the Physics and Mathematics of the Universe (WPI), UTIAS \\The University of Tokyo, Kashiwa, Chiba 277-8583, Japan}

\author{Philip Lu}\email{philiplu11@gmail.com}
\affiliation{Department of Physics and Astronomy, University of California, Los Angeles \\ Los Angeles, California, 90095-1547, USA}
\affiliation{Center for Theoretical Physics, Department of Physics and Astronomy, Seoul National University, Seoul 08826, Korea}

\author{Kohta Murase}\email{murase@psu.edu}
\affiliation{Department of Physics; Department of Astronomy and Astrophysics; Center for Multimessenger Astrophysics,
The Pennsylvania State University, University Park, Pennsylvania 16802, USA}
\affiliation{Center for Gravitational Physics, Yukawa Institute for Theoretical Physics, Kyoto, Kyoto 16802, Japan}

\author{Yoshiyuki Inoue}\email{yinoue@astro-osaka.jp}
\affiliation{Department of Earth and Space Science, Graduate School of Science, Osaka University, Toyonaka, Osaka 560-0043, Japan}
\affiliation{Interdisciplinary Theoretical \& Mathematical Science Program (iTHEMS), RIKEN, 2-1 Hirosawa, Saitama 351-0198, Japan} 
\affiliation{Kavli Institute for the Physics and Mathematics of the Universe (WPI), UTIAS \\The University of Tokyo, Kashiwa, Chiba 277-8583, Japan}

\author{Graciela B. Gelmini}\email{gelmini@physics.ucla.edu}
\affiliation{Department of Physics and Astronomy, University of California, Los Angeles \\ Los Angeles, California, 90095-1547, USA}
 
\date{\today}

\begin{abstract}
Primordial black holes (PBHs) formed in the early Universe constitute an attractive candidate for dark matter. Within the gaseous environment of the interstellar medium, PBHs with accretion disks naturally launch outflows such as winds and jets. PBHs with significant spin can sustain powerful relativistic jets and generate associated cocoons. Jets and winds can efficiently deposit their kinetic energies and heat the surrounding gas through shocks. Focusing on the Leo T dwarf galaxy, we demonstrate that these considerations can provide novel tests of PBHs over a significant $\sim 10^{-2} M_{\odot} - 10^6 M_{\odot}$ mass range, including the parameter space associated with gravitational wave observations by the LIGO and VIRGO Collaborations. 
Observing the morphology of emission could allow to distinguish between jet and wind contributions, and hence indirectly detect spinning PBHs. 
\end{abstract}
\maketitle

\section{Introduction}
Primordial black holes (PBHs) that could have formed in the early Universe prior to galaxies and stars can constitute a significant fraction of the dark matter (DM), can significantly affect the cosmological history and have been associated with a variety of observational signatures~(e.g.,~\cite{Zeldovich:1967,Hawking:1971ei,Carr:1974nx,Meszaros:1975ef,Carr:1975qj,GarciaBellido:1996qt,Kawasaki:1997ju,Khlopov:2008qy,Carr:2009jm,Frampton:2010sw,Fuller:2014rza,Bird:2016dcv,Kawasaki:2016pql,Inomata:2016rbd,Pi:2017gih,Inomata:2017okj,Garcia-Bellido:2017aan,Georg:2017mqk,Kocsis:2017yty,Ando:2017veq,Cotner:2016cvr,Cotner:2019ykd,Cotner:2018vug,Sasaki:2018dmp,Carr:2018rid,Kawasaki:2018daf,Takhistov:2017bpt,Takhistov:2017nmt,Flores:2020drq,Deng:2017uwc,Kusenko:2020pcg,Carr:2020gox,Green:2020jor,Takhistov:2020vxs}). Depending on the formation scenario, the mass of PBHs can span many orders of magnitude.

Particularly intriguing are PBHs in the stellar BH mass range of $\sim 10-10^{2}M_{\odot}$, which have been directly linked with the breakthrough observations of gravitational waves (GWs) by the LIGO and Virgo collaborations (LVC)~\cite{LIGOScientific:2016aoc}. Dozens of binary BH merger sources have already been detected~\cite{LIGOScientific:2020ibl}.
While uncertainties exist~(e.g.,~\cite{Jedamzik:2020ypm,Jedamzik:2020omx}), only a fraction $f_{\rm PBH} \lesssim \mathcal{O}(10^{-3})$ of PBHs contributing to the DM energy density, $f_{\rm PBH} = (\Omega_{\rm PBH}/\Omega_{\rm DM})$, is needed to account for the GW data~\cite{Bird:2016dcv,Clesse:2016vqa,Sasaki:2016jop,Wang:2016ana,Ali-Haimoud:2017rtz,Clesse:2017bsw,Raidal:2018bbj,Vaskonen:2019jpv,Hall:2020daa,DeLuca:2021wjr,Hutsi:2020sol,Wong:2020yig,Franciolini:2021tla}. 

The first detected intermediate mass BH, the GW190521 LVC event, with a total merger mass of $\sim 150 M_{\odot}$ that lies in the pair instability supernova mass gap~\cite{LIGOScientific:2020iuh}, has challenged conventional astrophysical interpretations. Further, GW candidate events consistent with solar mass BHs, lying in the lower mass gap region $\lesssim 3 M_{\odot}$, have drawn speculations about possible non-astrophysical origins ~(e.g.,~\cite{Capela:2013yf,Fuller:2017uyd,Bramante:2017ulk,Takhistov:2017bpt,Takhistov:2017nmt,Genolini:2020ejw,Kouvaris:2018wnh,Tsai:2020hpi,Takhistov:2020vxs,Dasgupta:2020mqg,Sasaki:2021iuc}).

While a variety of different constraints exist over the $\sim 1-10^4 M_{\odot}$ mass range for PBHs contributing significantly to the DM abundance~\cite{Macho:2000nvd,Monroy:2014,Ali-Haimoud:2016mbv,Inoue:2017csr,Poulin:2017bwe,Oguri:2017ock,Zoutendijk:2020,Lu:2020bmd,Serpico:2020ehh,Takhistov:2021aqx}, stellar mass and intermediate mass PBHs have often been considered as nonrotating (Schwarzschild) BHs~\cite{Chiba:2017rvs,DeLuca:2019buf,Mirbabayi:2019uph}.
Recently, outflow emission has been proposed as a new observable for studying accreting nonrotating PBHs~\cite{Lu:2020bmd,Takhistov:2021aqx}. 

PBHs can be formed with significant spin (Kerr BHs)~(e.g.,~\cite{Amendola:2017xhl,Flores:2020drq,Domenech:2021uyx, Harada:2016mhb,Kokubu:2018fxy,Cotner:2016cvr,Cotner:2019ykd,Cotner:2018vug,Flores:2021tmc}) as well as acquire spin via accretion~\cite{DeLuca:2020bjf} or hierarchical mergers~\cite{Fishbach:2017dwv}.~Aside from mass and charge, spin constitutes a fundamental conserved BH parameter.~Recent works focusing on small PBHs undergoing efficient Hawking evaporation have shown that spin can significantly affect observations~(e.g.,~\cite{Carr:2009jm,Arbey:2019vqx,Dong:2015yjs,Kuhnel:2019zbc,Arbey:2019jmj,Bai:2019zcd,Dasgupta:2019cae,Laha:2020vhg,Hooper:2020evu,Domenech:2021wkk}). 

In this work we analyze effects of outflow emissions from PBHs on the surrounding interstellar medium (ISM) gas for stellar and intermediate mass PBHs.
Spinning PBHs can support powerful relativistic jets, an important emission component that has been previously underexplored. We revisit emission from winds associated with accreting PBHs, which could be highly efficient. As we demonstrate, these combined effects allow for stringent tests of PBHs over a significant parameter space.
 
\begin{figure}[tb]
\begin{center}
\includegraphics[trim={0mm 0mm 0 0mm},clip,width=.47\textwidth]{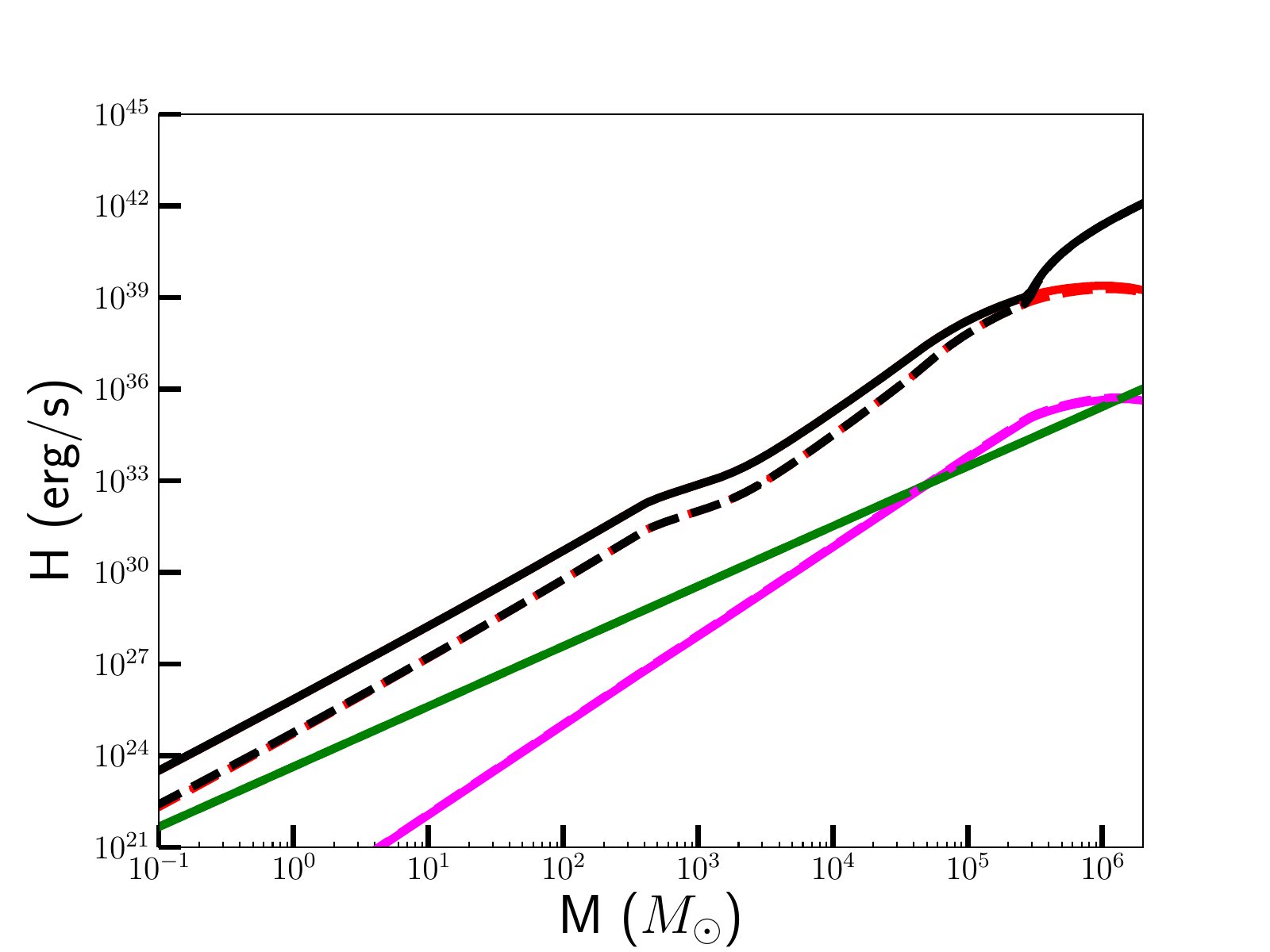}
\caption{ \label{fig:heatpbhlim}
Comparison between the spin $a = 0$ (dashed) and spin $a = 0.9$ (solid) cases of the effective gas heating from various contributions, bremsstrahlung (magenta), inverse Compton (red) and dynamical friction (green).
}
\end{center}
\end{figure}

\section{Accretion Disk Emission}
\label{sec:diskheat}
As PBHs traverse the ISM, they interact with the surrounding gas, depositing energy and heat. This has been recently analyzed in detail for Schwarzschild PBHs~\cite{Lu:2020bmd,Takhistov:2021aqx}. We discuss how the PBH spin affects accretion disk emission. We take the convention $c = 1$.

In the presence of non-negligible spin, the radius of the PBH innermost stable circular orbit (ISCO) that may mark the inner edge of the accretion disk is~\cite{Bardeen:1972fi}
\begin{equation}
r_{\rm min} = G M\left(3+Z_2 \mp \left[(3-Z_1)(3+Z_1+2Z_2)\right]^{1/2}\right)~,
\end{equation}
where
\begin{align}
&Z_1 \equiv 1+\Big(1-\chi^2\Big)^{1/3}\Big[\Big(1+\chi\Big)^{1/3}+\Big(1-\chi\Big)^{1/3}\Big]  \notag\\
&Z_2 \equiv \Big(3\chi^2+Z_1^2\Big)^{1/2}~.
\end{align}
Here $M$ is the PBH mass, $\chi = 2 a / R_s$, $a$ is spin Kerr parameter, and $R_s = 2 G M$ is the Schwarzschild radius, where natural units are adopted.
 
To estimate the change in PBH emission with spin, we focus on the radiatively inefficient accretion flows (RIAFs) that form when the accretion is sub-Eddington~\cite{Yuan:2014gma},
as relevant for our parameter space of interest~\cite{Lu:2020bmd,Takhistov:2021aqx}.
We follow the approximate analytic expressions of Ref.~\cite{Mahadevan:1996jf} to describe the complex multiblackbody RIAF spectrum, employing updated phenomenological input parameters describing emission, as in the analysis for Schwarzschild PBHs~\cite{Lu:2020bmd,Takhistov:2021aqx}. 
With respect to Refs.~\cite{Lu:2020bmd,Takhistov:2021aqx} the components are modified through non-trivial functions of temperature as well as other inputs dependent on $r_{\rm min}$.

Photons emitted from the accretion disk interact with and heat the surrounding ISM. The Bondi-Hoyle accretion rate is~\cite{1939PCPS...35..405H,1944MNRAS.104..273B,1952MNRAS.112..195B}
\begin{align} 
\label{eq:bondihoyle}
\dot{M} =&~ 4 \pi r_B^2 \tilde{v} \rho_a = \frac{4\pi G^2 M^2 n \mu m_p}{\Tilde{v}^3} \notag\\
\simeq &~ 2.3\times10^{31}~ \textrm{erg/s}~ \left(\frac{M}{1~M_\odot}\right)^2 \notag\\
&~\times \left(\frac{n}{0.07~\textrm{cm}^{-3}}\right) \left(\frac{\Tilde{v}}{10~\textrm{km/s}}\right)^{-3}~, 
\end{align}
where $r_B = 2 G M/\tilde{v}^2$ is the Bondi radius, $\rho_a$ is the ambient gas density, $\mu$ is the mean molecular weight, $n$ is the ISM gas number density, $m_p$ is the proton mass and $\tilde{v} \equiv (v^2+c_s^2)^{1/2}$. Here $v$ is the PBH velocity relative to the ISM gas and $c_s$ is the temperature-dependent sound speed in gas, which we take to be approximately $c_s \sim 10$~km/s~\cite{Inoue:2017csr}. 

Favorable systems, such as the dwarf galaxy Leo T, are rich in atomic hydrogen gas that strongly absorbs ionizing radiation, $E\gtrsim E_i=13.6\textrm{eV}$. However, atomic hydrogen is optically thin both below the $E_i$ threshold and for hard X-rays $E\gtrsim 1\textrm{keV}$. We consider the photo-ionization cross section \cite{Bethe1957,1990A&A...237..267B} to be $  \sigma(E) = \sigma_0 y^{-\frac{3}{2}}\left(1+y^{\frac{1}{2}}\right)^{-4}$,
where $y=E/E_0$, $E_0 = 1/2 E_i$ and $\sigma_0 = 
6.06\times10^{-16}\textrm{ cm}^2$. For energies above $30\textrm{ eV}$, we use the attenuation length data from Fig.~(32.16) of Ref.~\cite{Olive_2014}, which includes the Thompson (Compton) scattering cross-section that is dominant for high energies. 
The optical depth is $\tau (n, E) = \sigma(E) n r_{\rm sys}$, where $r_{\rm sys}$ is the characteristic size of the system. The heating rate is then a function of both the frequency-dependent luminosity $L_\nu$ and the optical depth, $\tau$,
\begin{equation}
\label{eq:heatingphotons}
H_{\rm pe}(M,n,v)=\int_{E_i}^{E_\textrm{max}}L_\nu (M,n,v) f_h \left(1-e^{-\tau}\right)d\nu~,
\end{equation}
with an additional factor of $f_h\sim 1/3$~\cite{2010MNRAS.404.1869F} for the fraction of energy loss deposited as heat. We integrate Eq.~\eqref{eq:heatingphotons} up to the electron temperature $E_\textrm{max}\sim {\rm few} \times T_e$, above which the emission falls off exponentially. 
 
As shown by the semi-analytic modeling~\citep{Mahadevan:1996jf}, the electron temperature in the accretion disk is set by the balance of heating and emission processes. Both viscous heating, which is dominant at low accretion rates, and ion-electron collisional heating for higher accretion rates scale approximately inversely with the ISCO radius $r_\textrm{min}$. We neglect the effects of collisionless heating in our treatment. Thus, assuming the disk and BH are aligned, higher PBH spins result in $r_\textrm{min}$ moving inwards and generally increase the electron temperature of the disk plasma. In turn, increased emission from the three counterbalancing processes: synchrotron, inverse Compton (IC), and bremsstrahlung, resolves the temperature to a new higher equilibrium. 

In Fig.~\ref{fig:heatpbhlim}, we display contributions to effective heating for Schwarzschild as well as spinning ($a = 0.9$) PBHs. 
Regarding heat deposition by emission from the innermost region around the ISCO, we find that spinning PBHs yield an order of magnitude increase in the IC contribution, but not the synchotron (not displayed) or the bremsstrahlung emissions. 
The synchrotron and IC luminosity of RIAFs have $L_{\rm syn}\propto T_e^7$ and $T_{e}^{6+\alpha_{\rm IC}}$ for $\alpha_{\rm IC}\lesssim1$~(e.g.,~\cite{Kimura:2020thg}), so the temperature dependence is stronger for the synchrotron in general.
For the bremsstrahlung spectrum, increasing the electron temperature does not significantly increase the luminosity throughout the spectrum, but rather further extends it to higher frequencies. However, our hydrogen-rich environments of interest such as the Leo T dwarf galaxy are optically thin to Compton scattering at these high energies, so the extended spectrum does not increase the deposited energy and heating. 

Similarly, although the synchotron luminosity is often dominant in RIAF disks, synchotron radiation predominantly emits in radio and infrared frequencies that interact negligibly with atomic hydrogen. Hence, we do not consider this contribution. The higher temperature of spinning black holes does increase the synchrotron emission and also shifts the synchotron peak to higher frequencies, closer to the ionization threshold of $13.6\textrm{ eV}$. Photons from this synchotron peak provide targets for the IC scattering. Thus, the augmented synchotron peak nearer to the ionization, along with more efficient IC upscattering at higher temperatures, significantly boosts the quantity of strongly absorbed ionizing IC photons, leading to a non-linear rise in the heating rate. We note that throughout we consider that emission follows a single zone model. However, significant uncertainties exist and other models are also possible. 
  
We also display in Fig.~\ref{fig:heatpbhlim} dynamical friction, with heat generated by dynamical friction force
$F_\textrm{dyn} = -4\pi G^2 M^2 \rho I/v^2 $. Here
$I$ is a geometrical integral that is only weakly affected by the ISCO/min radius through an additive $\ln\left(r_\textrm{max}/r_\textrm{min}\right)$ contribution when
$v/c_s>1$. Hence, the inclusion of PBH spin does not significantly affect this energy deposit component.
 
\begin{figure}[tb]
\begin{center}
\includegraphics[trim={15mm 0mm 0mm 0mm},clip,width=.49\textwidth]{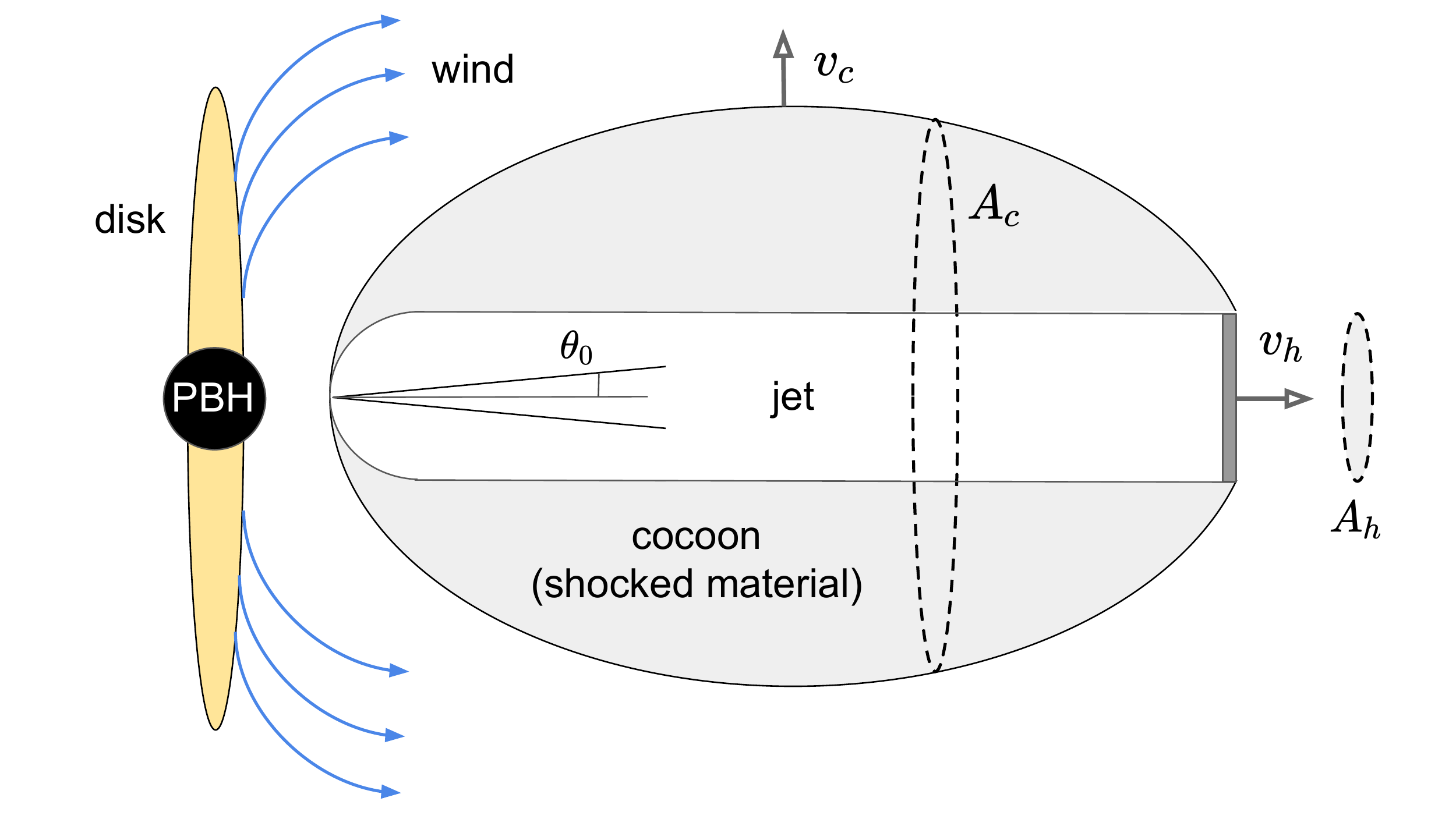}
\caption{\label{fig:outflow} 
Schematic diagram for PBH jet with cocoon, associated with a spinning PBH, and wind outflow emission. The cocoon shock expands into the ISM with speeds $v_h$ along the jet-axis and $v_c$ in orthogonal directions. Mean cocoon cross-sectional area $A_c$, area of the tip bow shock $A_h$ as well as jet opening angle $\theta_0$ are shown. 
}
\end{center}
\end{figure}

\section{Jets and Cocoons}
\label{sec:jets}
In addition to emission from the accretion disk, spinning astrophysical BHs can naturally power collimated relativistic jets~\footnote{Under specific circumstances, the formation of jets via other channels, such as neutrino-antineutrino annihilation~(e.g.,~\cite{Janka:1999qu}) or magnetocentrifugal force through the disk magnetic field (Blandford-Payne)~\cite{Blandford:1982} may be possible.} via the electromagnetic extraction of rotational energy from a BH through the Blandford-Znajek (BZ) mechanism~\cite{Blandford:1977ds}. We now discuss the emission associated with jets and related cocoons for highly spinning ($a > 0.5$) PBHs traversing the ISM.

With the strong magnetic fields relevant for jet launching, numerical simulations indicate that ``magnetically-arrested'' disks (MADs) can be expected to form~\cite{Narayan:2003by,Tchekhovskoy:2011}, allowing for a highly efficient engine powering the jet with a luminosity
\begin{equation}
L_{\rm j}= \epsilon_{j}\dot{M}_{\rm acc}~,
\end{equation}
where $\epsilon_j \sim \mathcal{O}(1)$ is the jet efficiency factor~\cite{Narayan2021arXiv210812380N} and $\dot{M}_{\rm acc}$ is the disk accretion rate given by Eq.~\eqref{eq:bondihoyle}.
Such a scenario is consistent with the recently imaged ring of the M87 galaxy central BH~\cite{EventHorizonTelescope:2019pgp} as well as the analysis of blazar spectral energy distribution~\cite{Ghisellini2014Natur.515..376G}~\footnote{However, extended jet studies indicate $\epsilon_j \sim \mathcal{O}(10^{-2})$~\cite{vanVelzen2013A&A...557L...7V, Pjanka2017MNRAS.465.3506P,Inoue2017ApJ...840...46I}.}. 

\begin{figure}[tb]
\begin{center}
\includegraphics[trim={0mm 0mm 0 0},clip,width=.45\textwidth]{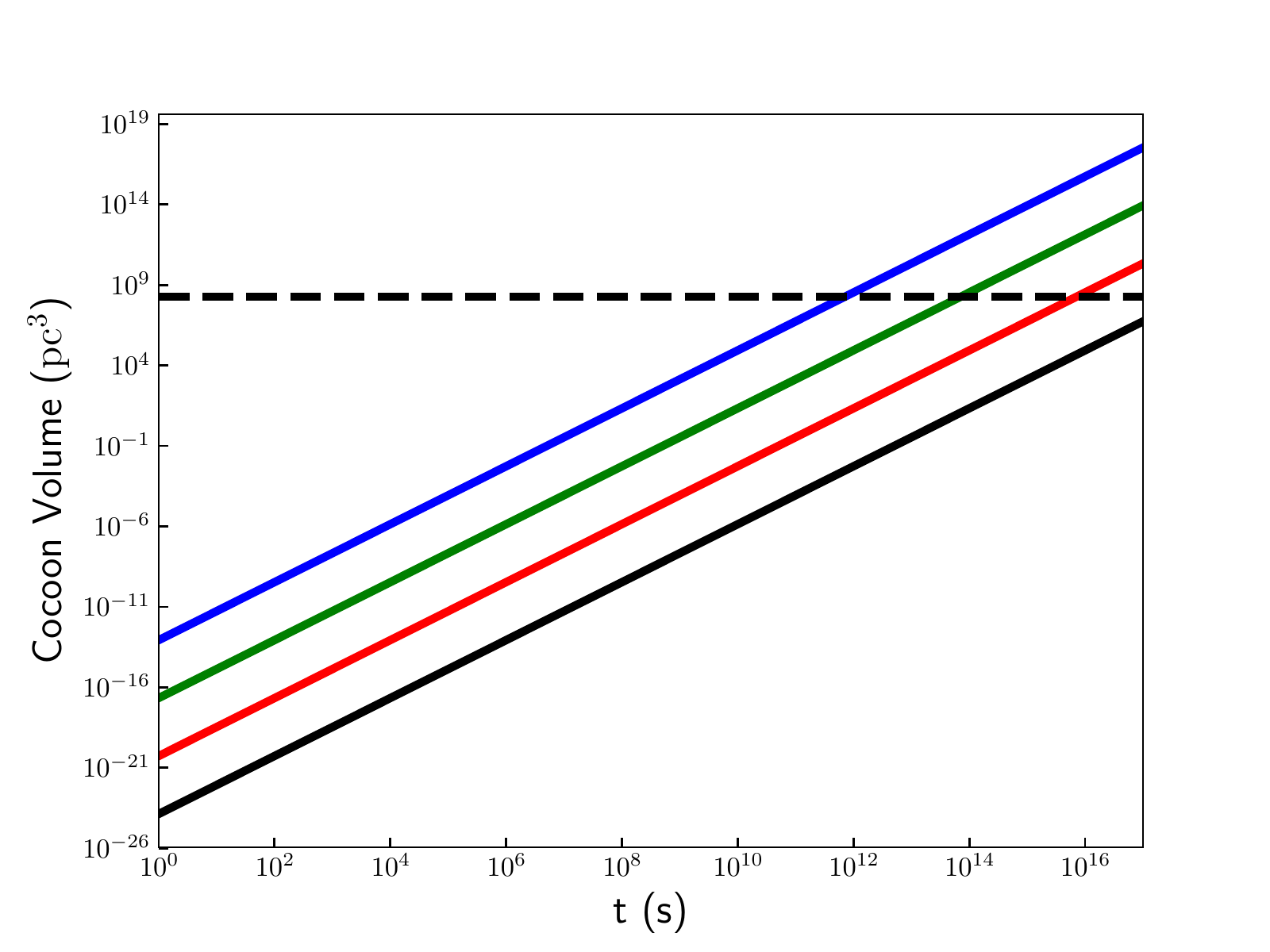}
\caption{\label{fig:cocoon} 
Cocoon volume as a function of time for a gas system such as the Leo T dwarf galaxy. Results for PBHs of masses M=$10^{-3} M_\odot$ (black), $1 M_\odot$ (red), $10^{3} M_\odot$ (green), and $10^{6} M_\odot$ (blue) are shown. For reference, the black dashed line indicates the total volume of the inner region of Leo T, $1.8\times10^{8} \textrm{pc}^3$.
}
\end{center}
\end{figure}

The jet interacts with external material, forming a pair of forward and reverse shocks. Then a significant $\mathcal{O}(1)$ portion of the jet power is deposited into an expanding ``cocoon'' consisting of the shocked material~\cite{Begelman:1989}. In Fig.~\ref{fig:cocoon} we schematically depict the cocoon structure.

A fraction of the cocoon's energy is radiated away as thermal bremsstrahlung emission, predominantly from young cocoons as $L_{\rm brem} \sim t^{-1}$~\cite{Kino:2009kb}. For the scenario we consider, the cocoons are long lived and they eventually deliver most of their energy efficiently to the surrounding ISM (i.e., the associated system heating by the outflow is $H_\textrm{out}\sim\dot{M}_{\rm in}$). 
To quantify the uncertainties, we consider a range from $H_\textrm{out}=\dot{M}_{\rm acc}$ 
to $H_\textrm{out}=5 \times 10^{-3} \dot{M}_{\rm acc}$. 
The former value is motivated by jets associated with MADs. 
The latter value, which is more conservative, is motivated by the AGN feedback consideration for the wind although it may also be achieved by inefficient jets.

The cocoon can be analytically characterized using only the jet luminosity $L_j$, the density of the ambient gas medium $\rho_a$ and the jet opening angle $\theta_0\sim$few degrees~\cite{Bromberg:2011}.
Using Eq. (2) and (3) of Ref.~\cite{Begelman:1989} and Eq. (12)~of Ref.~\cite{Bromberg:2011}, and considering a relativistic jet propagation speed of $v_j \simeq 1$, we obtain the cocoon sideways expansion speed
\begin{align} \label{eq:vc}
v_c \approx &~ \left(\frac{L_j \theta_0}{2 \rho_a t^2}\right)^{1/5} \notag \\
    \sim &~ 10^{3}~ \textrm{km/s}~\, \epsilon_j^{1/5} \left(\frac{M}{1~M_\odot}\right)^{2/5} \left(\frac{\Tilde{v}}{10\textrm{ km/s}}\right)^{-3/5} \notag \\ &~\times  \left(\frac{\theta_0}{0.1}\right)^{1/5}\left(\frac{t}{1\textrm{ yr}}\right)^{-2/5}~.
\end{align}
Using Eq.~\eqref{eq:vc} with Eq.~(2) of Ref.~\cite{Begelman:1989}, and considering that the jet's mean cross-sectional area is $A_c \simeq v_c^2 t^2$, the speed of cocoon expansion along the jet-axis is
\begin{align}
v_h \approx &~ \frac{L_j t^2}{\rho_a A_c^2} \approx \Big(\frac{2^{4} L_j}{\rho_a t^{2}\theta_0^{4}}\Big)^{1/5} \notag \\
    \sim &~ 2\times10^{4}~ \textrm{km/s}~\, \epsilon_j^{1/5}    \left(\frac{M}{1~M_\odot}\right)^{2/5} \left(\frac{\Tilde{v}}{10\textrm{ km/s}}\right)^{-3/5} \notag \\
    &~ \times \left(\frac{\theta_0}{0.1}\right)^{-4/5}\left(\frac{t}{1\textrm{ yr}}\right)^{-2/5}~.
\end{align}
The resulting cocoon volume is then
\begin{align}
V_c \approx &~ v_c^2 v_h t^3  \approx 
    \Big(\frac{2^{2} t^{9} L_j^{3}}{\rho_a^{3}\theta_0^{2}}\Big)^{1/5} \notag \\
    \simeq &~ 2.1\times10^{-8}\textrm{ pc}^3 \epsilon_j^{3/5}   \left(\frac{M}{1~M_\odot}\right)^{6/5} \left(\frac{\Tilde{v}}{10\textrm{ km/s}}\right)^{-9/5} \notag \\
    & ~ \times \left(\frac{\theta_0}{0.1}\right)^{-2/5}\left(\frac{t}{1\textrm{ yr}}\right)^{9/5}~.
\end{align} 

The time $t_b$ it takes for the expanding cocoon width to reach the Bondi radius is
\begin{align}
\label{eq:tbondi}
t_b \approx &~ \left(\frac{6GM}{5\tilde{v}^{2}}\right)^{5/3}\left(\frac{2\rho_a}{L_j \theta_0}\right)^{1/3} \notag\\ 
    \simeq &~ 6.9\times10^{-3}\textrm{ yr}~\epsilon_j^{-1/3}\left(\frac{M}{1~M_\odot}\right) \notag\\
    &\times \left(\frac{\Tilde{v}}{10\textrm{ km/s}}\right)^{-7/3}\left(\frac{\theta_0}{0.1}\right)^{-1/3}~.
\end{align}

The number of PBHs of mass $M$ contributing a fraction $f_{\rm DM}$ of the DM abundance within a system of radius $r_{\rm sys}$ and uniform DM density $\rho_{\rm DM}$ is
\begin{align} 
N_{\rm PBH} =&~ \frac{4}{3} \pi r_{\rm sys}^3 \Big(\frac{\rho_{\rm DM} f_{\rm PBH}}{M}\Big) \notag\\ 
     \simeq &~ 8.3\times10^{6} ~f_{\rm PBH} \left(\frac{r_{\rm sys}}{350\textrm{ pc}}\right)^3 \notag\\
     &~\times \left(\frac{\rho_{\rm DM}}{1.75\textrm{ GeV}/\textrm{cm}^3}\right)\left(\frac{M}{1~M_\odot}\right)^{-1}~. 
\end{align}
Solving the equation $V_{\rm sys} = N_{\rm PBH} V_c(t_{\rm fill})$, the time $t_{\rm fill}$ it takes for the volume of PBH cocoons to fill up the volume $V_{\rm sys} = (4/3) \pi r_{\rm sys}^3$ of a system of gas is 
\begin{align}  \label{eq:tfill}
t_{\rm fill} \approx &~ \left(\frac{M}{\rho_{\rm DM}f_{\rm PBH}}\right)^{5/9}\frac{\tilde{v}\theta_0^{2/9}}{2^{2/9}(4\pi G^2M^2)^{1/3}} \notag \\
    \sim &~ 10^{5} \textrm{ yr} ~ f_{\rm PBH}^{-5/9} \left(\frac{M}{1~M_\odot}\right)^{-1/9}\left(\frac{\Tilde{v}}{10\textrm{ km/s}}\right) \notag\\ 
    &~\times \left(\frac{\theta_0}{0.1}\right)^{2/9} \left(\frac{\rho_{\rm DM}}{1.75\textrm{ GeV}/\textrm{cm}^{-3}}\right)^{-5/9}~. 
\end{align}
We note that Eq.~\eqref{eq:tfill} is independent of the system size.

Comparing Eq.~\eqref{eq:tfill} with Eq.~\eqref{eq:tbondi} allows to define the condition $ t_b \gg t_{\rm fill}$ such that effects of feedback will not have any potential impact on PBH accretion during timescales $t_{\rm fill}$ associated with the volume of PBH cocoons filling the system of interest, which requires
\begin{align}
\tilde{v} \ll&~6 \times 10^{-2}~\textrm{km/s}~ \epsilon_j^{-1/10} f_{\rm PBH}^{1/6} \Big(\frac{\theta_0}{0.1}\Big)^{-1/6} \notag\\ &\times \Big(\frac{M}{1~M_{\odot}}\Big)^{1/3} \left(\frac{\rho_{\rm DM}}{1.75\textrm{ GeV}/\textrm{cm}^{-3}}\right)^{1/6}~.
\end{align}
This condition is difficult to satisfy except e.g. for a small jet opening angle $\theta_0$. Hence, the complication effects associated with feedback could be in principle of some relevance. However, it is reasonable to expect that efficient accretion will still persist, e.g. in the equatorial plane.

\section{Winds}
Aside from collimated jets associated with spinning PBHs, both spinning and nonspinning PBHs can naturally form sub-relativistic outflows of ionized gas through a variety of mechanisms~(e.g.,~\cite{Begelman:1983,Krolik:2001,Proga:2004jf,Fukumura:2009gn}). The significance of such disk-driven winds has been recently highlighted for nonrotating PBHs~\cite{Lu:2020bmd,Takhistov:2021aqx}. There, the winds were treated employing a phenomenological self-similar model~\cite{Yuan:2014gma} and their energy deposit was estimated from stopping power considerations. 
However, the description of winds is uncertain and the winds should rather deposit a sizable amount of energy into surrounding material via shock heating, similarly to cocoons. More so, the wind energy deposit can be efficient (e.g.,~\cite{Liu:2017bjr,DiMatteo2005Natur.433..604D}). 
Hence, here we consider a simplified phenomenological treatment, parametrizing the wind power analogously to that of jets and cocoons with wind luminosity $L_w$ given by
\begin{equation}
L_w = \epsilon_w \dot{M}_{\rm out}~,
\end{equation}
where $\epsilon_w$ is the wind efficiency factor. 

The ratio of power efficiencies of jets and winds is a matter of ongoing debate (see e.g.,~\cite{Mehdipour:2019cks} for discussion of possible wind-jet connections). In our subsequent discussion of gas heating, we assume different efficiency contributions of jets and winds. 
For the wind, motivated by the AGN feedback consideration, we adopt $H_{\rm out}=5 \times 10^{-3} \dot{M}_{\rm acc}$, which implies that $\sim5$\% of the quasar bolometric luminosity (that is typically $\sim 0.1\dot{M}_{\rm acc}$)
is dissipated for the ISM heating~\cite{DiMatteo2005Natur.433..604D}. 
We do not distinguish between the individual wind and jet contributions to gas heating. However, we expect that the jet contribution is more significant, e.g. in case of MADs~\cite{Tchekhovskoy:2011,Tchekhovskoy:2015}.
Observationally, jets can be distinguished from winds by analysis of the emission morphology and the detection of ``hot spots'' associated with strong shocks where jets terminate~\cite{Blandford:2018iot}. 
Jet signatures from isolated BHs have been studied in Ref.~\cite{2017MNRAS.470.3332I}.

\section{Gas Heating}
\label{sec:gasheating}

\begin{figure}[tb]
\begin{center}
\includegraphics[trim={0mm 0mm 0 0mm},clip,width=.50\textwidth]{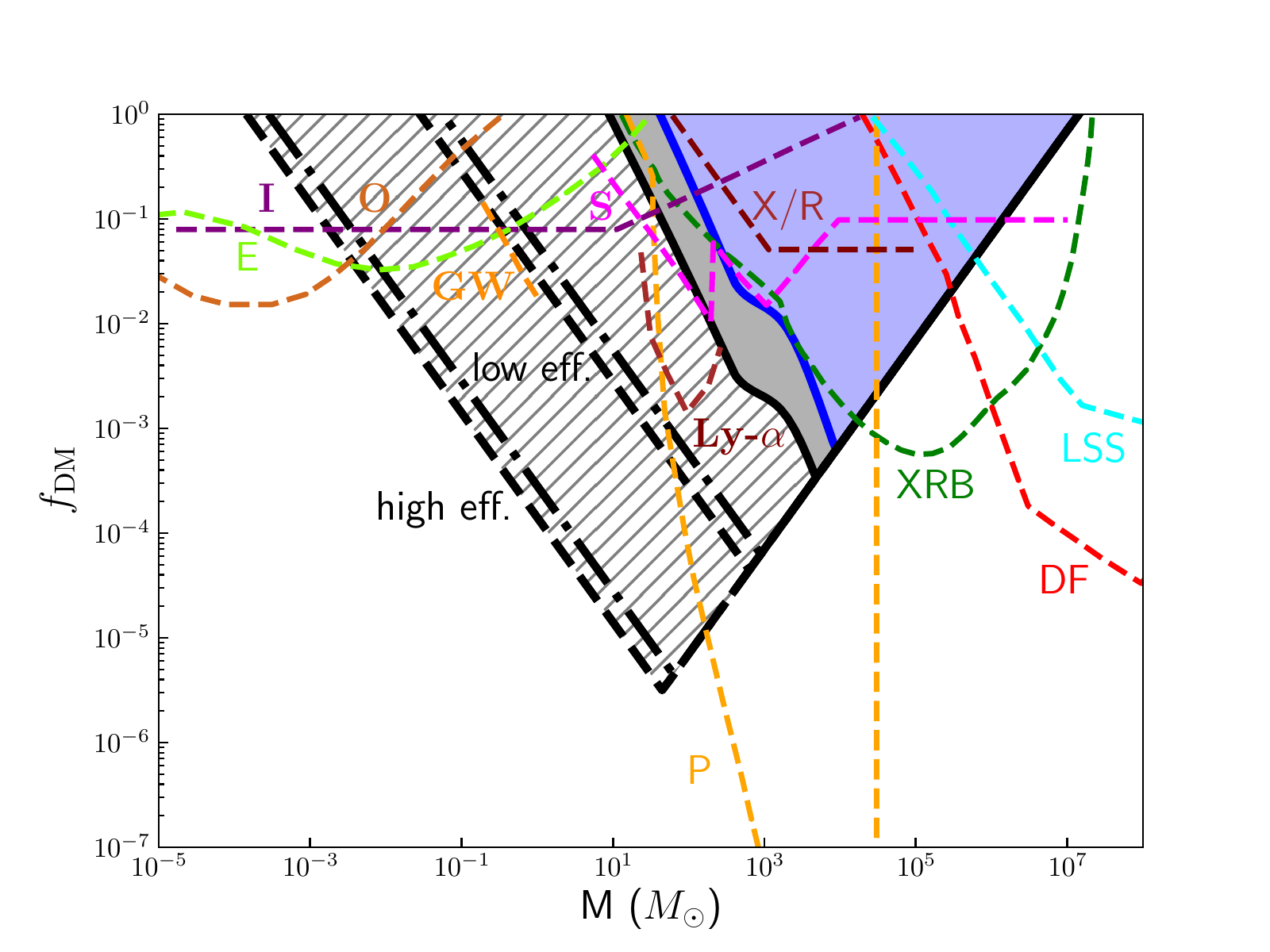}
\caption{ \label{fig:pbhconstraints} 
Constraints from PBH gas heating of the Leo T dwarf galaxy. Effects of jets and winds are shown with black hatching over a broad range between $H_{\rm out} = \dot{M}_{\rm acc}$ (``high eff.'') and $H_{\rm out} = 5\times10^{-3}\dot{M}_{\rm acc}$ (``low eff.''), assuming all PBHs form efficient jets and winds (black dashed) or that only 50\% do (black dot-dashed). Gas heating contributions from disk emission and dynamical friction are shown in blue and of spinning PBHs in black (see text).  
These constraints are bounded by the PBH incredulity limit (positive slope line). Other existing constraints (see Ref.~\citep{Carr:2020gox})  are shown by dashed lines including Icarus~\cite{2018PhRvD..97b3518O} (I) caustic crossing in purple, Planck~\cite{Ali-Haimoud:2016mbv,Serpico:2020ehh} (P) in yellow, X-ray binaries~\cite{Inoue:2017csr} (XRB) in green, dynamical friction of halo objects (DF) in red, Lyman-$\alpha$~\cite{2019PhRvL.123g1102M} (Ly-$\alpha$) in maroon, combined bounds from the survival of astrophysical systems in Eridanus II~\cite{Brandt:2016aco}, Segue 1~\cite{2017PhRvL.119d1102K}, and disruption of wide binaries~\cite{2014ApJ...790..159M} (S) shown in magenta, large scale structure~\cite{Carr:2018rid} (LSS) in cyan, X-ray/radio~\cite{Ricotti:2007au} (X/R) in brown, and gravitational waves (GWs) in green~\cite{LIGOScientific:2019kan}.
}
\end{center}
\end{figure}

The deposition of energy by PBHs into the ISM heats the surrounding gas. Analyses of stellar and intermediate mass Schwarzschild PBHs~\cite{Lu:2020bmd,Takhistov:2021aqx} as well as evaporating PBHs~\cite{Laha:2020vhg,Kim:2020ngi} demonstrated that considerations based on balancing gas heating and cooling within a target system is a powerful method for exploring and constraining PBHs\footnote{Our approach to set the limits is similar to that used for particle DM~\citep{Bhoonah:2018wmw,Farrar:2019qrv,Wadekar:2019xnf}, but the heating mechanisms and the preferred gas systems are different in our case.}, considering particularly favorable environments such as the DM-rich Leo T dwarf galaxy. Using the results obtained above, we now discuss gas heating in Leo T due to PBHs. Our analysis can be readily applied to other systems.

PBHs heat the gas system by a total amount $H_{\rm tot} = N_\textrm{PBH} H(M)= f_\textrm{PBH}\rho_\textrm{DM}V_{\rm sys} H(M)/M$, where $H(M)$ is the average heat generated
by one PBH of mass $M$ through the different contributions. Requiring the total heating not to exceed the total cooling $\dot{C} V_{\rm sys}$ of the system, where $V_{\rm sys}$ is the relevant volume of interest and $\dot{C}$ is the cooling rate, the PBH DM abundance $f_{\rm PBH}$ is constrained by
\begin{align}
\label{eq:genbound}
f_\textrm{PBH} <&~ f_\textrm{bound} =\frac{M\dot{C}}{\rho_\textrm{DM}H(M)} \\
& \simeq  6\times10^{-5} \left(\frac{M}{1~ M_\odot}\right)\left(\frac{\dot{C}}{2.28\times10^{-30}~\textrm{erg/(cm$^3$ s)}}\right) \notag\\ &~~~\times \left(\frac{\rho_{\rm DM}}{1.75\textrm{ GeV}/\textrm{cm}^{-3}}\right)^{-1} \left(\frac{H(M)}{2.3\times10^{31}~\textrm{erg/s}}\right)^{-1}, \notag
\end{align}
where the inserted quantities are characteristic of Leo T with $H(M)$ being efficient jet heating.

We consider gas heating  due to the combined contributions of accretion disk emission Eq.~\eqref{eq:heatingphotons}, dynamical friction $F_{\rm dyn}$, and outflows associated with jets and winds whose heating rate is $H_{\rm out} \simeq L_j + L_w$.
For the system, we limit our considerations to the central $r\lesssim 350 \textrm{ pc}$ region of Leo T, which is dominated by atomic hydrogen\footnote{The gas outside is expected to be ionized~\citep{2013ApJ...777..119F}, resulting in efficient cooling~\citep{Smith:2016hsc}.}~\citep{2013ApJ...777..119F}. For the varying central gas density we take the approximate constant value $n=0.07\textrm{ cm}^{-3}$~\cite{2013ApJ...777..119F} and $\rho_{\rm DM} = 1.75 \textrm{ GeV}/\textrm{cm}^{3}$. Focusing on the dominant gas component, we consider a velocity dispersion of $\sigma_g=6.9~ \textrm{km}/\textrm{s}$ and a temperature $T\simeq 6000$~K~\citep{2008MNRAS.384..535R,2013ApJ...777..119F}. Since the DM is expected to have the same velocity dispersion as the gas, we set $\sigma_v=\sigma_g$. From the adiabatic formula with input temperature of $T\simeq 6000$~K we determine the sound speed to be $c_s=9\textrm{ km}/\textrm{s}$. Combining the radius and number density, we obtain the column density of hydrogen gas in the central region of Leo T is $n r_\textrm{sys} = 7.56\times10^{19}\textrm{ cm}^{-2}$. For the gas metallicity we consider that it approximately follows the stellar one $\textrm{[Fe/H]} \simeq -2$ estimated from the stellar spectra~\citep{2008ApJ...685L..43K}, which is accurate up to a factor of few.

For the system cooling rate, we employ the approximate results of Ref.~\cite{Wadekar:2019xnf}. For hydrogen gas it is
\begin{equation}
\label{eq:coolinggen}
\dot{C} = \left(\frac{n}{1~{\rm cm}^{-3}}\right)^2 10^{[\textrm{Fe/H}]}\Lambda(T)~ \frac{\rm erg}{{\rm cm}^3{\rm s}}~,
\end{equation}
where [Fe/H]$~\equiv \log_{10}(n_{\rm Fe}/n_{\rm H})_{\rm gas} - \log_{10}(n_{\rm Fe}/n_{\rm H})_{\rm Sun}$ is the metallicity, and $\Lambda(T) \propto 10^{[{\rm Fe/H}]}$ is the cooling function. Fitting numerically results of the chemical network library~\cite{Smith:2016hsc}, one can obtain that $\Lambda(T) = 2.51\times10^{-28}(T/{\rm K})^{0.6}$ is valid for $300~\text{K} < T < 8000~\text{K}$~\citep{Wadekar:2019xnf}. For Leo T, $10^{{\rm [Fe/H]}} \simeq 10^{-2}$.

Conservatively, we neglect possible additional heating contributions from natural sources such as stellar radiation. Further, we ensure that the system is stable by requiring that its lifetime $\tau_\textrm{sys}$ significantly exceeds the thermalization timescale $t_{\rm therm}$, given by
\begin{align} \label{eq:therm}
t_{\rm therm} =&~ \frac{3}{2} \frac{nkT}{\dot{C}} \\
    \simeq &~ 1.2\times10^{9}\textrm{ yr} \left(\frac{T}{6000~{\rm K}}\right) \notag\\
    &\times  \left(\frac{n}{0.07~{\rm cm}^{-3}}\right)\left(\frac{\dot{C}}{2.28\times10^{-30}~\textrm{erg}/({\rm cm}^3~ {\rm s})}\right)^{-1}~, \notag
\end{align}
where $k$ is the Boltzmann constant.
A time for cocoons to fill the system shorter than $t_{\rm therm}$ ensures a continuous heating/cooling process. Comparing Eq.~\eqref{eq:tfill} to Eq.~\eqref{eq:therm}, $t_{\rm fill} \ll t_{\rm therm}$ implies
\begin{align}
f_{\rm PBH} >&~5.2\times10^{-8} \left(\frac{\tilde{v}}{10~\textrm{km/s}}\right)^{9/5} \left(\frac{M}{1~M_\odot}\right)^{-1/5} \notag\\ &\times \left(\frac{\rho_{\rm DM}}{1.75~{\rm GeV}/{\rm cm}^3}\right)^{-1}\left(\frac{n}{0.07~{\rm cm}^{-3}}\right)^{-9/5}\theta_0^{2/5} \notag\\ & \times \left(\frac{\dot{C}}{2.21\times10^{-30}~\textrm{erg/(cm$^3$ s)}}\right)^{9/5} \left(\frac{T}{6000~{\rm K}}\right)^{-9/5}~.
\end{align}
This condition does not impact our limits because they are more stringent. 

PBH effects are relevant only if statistically there is at least one PBH within the considered system, i.e. if $f_{\rm PBH}\, \rho_{\rm DM} (4\pi r_{\rm sys}^3/3)/M > 1$. This establishes the incredulity limit. Hence, we restrict our constraints to
\begin{equation}
\label{eq:incredlimit}
f_\textrm{PBH} > \frac{3M}{4\pi r_\textrm{sys}^3 \rho_\textrm{DM}}~.
\end{equation}

In Fig.~\ref{fig:pbhconstraints} we display the resulting limits from gas heating in Leo T on PBHs contributing to the DM, along with other existing constraints. The  constraints we found  due to PBH outflows (jets or winds), are shown in the hatched black regions. The dashed lines correspond to different contributions $H_{\rm out}$ to the heating rate, $H_{\rm out} = \dot{M}_{\rm acc}$ (``high eff.'') corresponds to jets associated with highly efficient magnetically arrested disks, while the conservative $H_{\rm out} = 5\times10^{-3}\dot{M}_{\rm acc}$ (``low eff.'') is suggested by AGN feedback for disk-winds. All intermediate $H_{\rm out}$ are also possible.  Jets and winds can act as sensitive probes of PBH parameter space and stringent constraints can be placed by PBH jets.

~\newline
\section{Conclusions}
\label{sec:conclusions}
  
PBHs formed in the early Universe can contribute a fraction or all of the DM abundance and have been directly linked with GW and other observations.
We have studied for the first time the energy deposition and contributions to ISM gas heating from jets with cocoons associated with spinning PBHs as well as outflowing winds. We have demonstrated that these combined outflow effects can act as sensitive probes of PBHs over orders of magnitude in mass range, from $\sim 10^{-2} M_{\odot}$ to $10^6 M_{\odot}$. This range is particularly of interest for LIGO/VIRGO GW events. The robustness and strength of our results is further established for spinning PBHs, which can sustain highly efficient relativistic jets in addition to winds as a separate powerful emission source. Contributions of jets and winds can be discriminated by analysis of the emission morphology as well as the detection of jet hot spots, allowing the potential for indirect insights into the spin distribution (and hence formation models) of PBHs. 

\section*{Acknowledgments}
We thank Kunihito Ioka for discussion. 
The work of G.B.G. and P.L. was supported in part by the U.S. Department of Energy (DOE) Grant No.~DE-SC0009937. P.L is also supported by Grant Korea NRF-2019R1C1C1010050. 
The work of K.M. is supported by the NSF Grant No.~AST-1908689, No.~AST-2108466 and No.~AST-2108467, and KAKENHI No.~20H01901 and No.~20H05852. 
The work of Y.I. is supported by JSPS KAKENHI grant No.~JP18H05458, JP19K14772, program of Leading Initiative for Excellent Young Researchers, MEXT, Japan, and RIKEN iTHEMS Program. V.T., and Y.I. are also supported by the World Premier International Research Center Initiative (WPI), MEXT, Japan. This work was performed in part at the Aspen Center for Physics, which is supported by National Science Foundation grant PHY-1607611. 

\bibliographystyle{bibi}
\bibliography{bibliography}
 
\end{document}